\documentstyle[twocolumn,aps]{revtex}
\begin{document}
\pagestyle{empty}
\title{Manganites at Low Temperatures and Light Doping:
Band Approach and Percolation}
\author{Lev P. Gor'kov$^{1}$ and Vladimir Z. Kresin$^2$}
\address{$^1$National High Magnetic Field
Laboratory, Florida State University, Tallahassee, FL 32310, U.S.A. and
L. D. Landau Institute for Theoretical Physics, Russia} 
\address{$^2$Lawrence Berkeley Laboratory,
University of California, Berkeley, 
CA 94720 }
\maketitle
\centerline {(Submitted \hspace{.5in} )}

\begin{abstract}
A tight-band model is
employed for the 
$e_{2g}$-orbitals in manganites.   It is shown that a large intraatomic
Hund's coupling, $J_H$ and the resulting 
double-exchange mechanism lead to antiferromagnetic ordering along one of
the cubic axis stabilized by the cooperative JT effect which further
decreases the band energy of electrons.  As a result, LaMnO$_3$ is a band
insulator built of 2D ferromagnetic layers.  
The critical concentration $x_c\simeq 0.16$, for onset of ferromagnetic
and metallic behavior at low temperatures in La$_{1-x}$Sr$_x$MnO$_3$ and
the phase 
transition are treated in the
percolation approach.

\vspace{.4cm}

\end{abstract}

\vspace{.25in}

Current efforts in studies of the ``colossal magnetoresistance''
(CMR) in the manganites, R$_{1-x}$B$_x$MnO$_3$ (usually, R=La, B=Ca, Sr;
see e.g. 
[1-3] for a review and references) are, naturally, focused on phenomena
occurring in the vicinity of the metal-insulator transition temperature,
T*, which for $x=0.2-0.3$, reaches the room
temperature range.  There are no doubts that both the Jahn-Teller (JT)
distortions and Zener's double exchange (DE) mechanism [4,5] are two key
ingredients of CMR.

The transition between the paramagnetic and
conducting ferromagnetic phases is treated in [6] as a
localization-delocalization ``crossover'' driven by thermal fluctuations of
the local JT-modes.  In this paper we focus on the analysis of phenomena
occurring at low T. In this 
region the approach should be different.  Indeed,the two
end members of the series, LaMnO$_3$ and CaMnO$_3$,are both antiferromagnetic
(AF) insulators.  ``Doping'' of LaMnO$_3$ by divalent atoms results in metallic
conductivity only above some critical concentration, $x_c\simeq 0.16$.

Low temperature phases of manganites are habitually interpreted in
terms of localized orbitals [7, 8], more often that not in the modified
Hubbard model (see. e.g. [9]).  Although it is widely accepted that
electron-electron 
correlations play the key role in low temperature behavior of
manganites, we show that their major properties can be understood within
the framework of the band approach, as well, or in percolation terms.   

The following
facts have to be explained: 1) the insulating state of the parent LaMnO$_3$;
2) the A-type of antiferromagnetic ordering (alternating ferromagnetic
planes); 3) the small value of the N\'{e}el temperature, $T_N\sim 150$K
in the low 
x-range, while the structural changes occur at about 900K; 4) the
localization of ``holes'' introduced into the Mn$^{3+}$-subsystem at
$x<x_c$; 5) the 
threshold concentration, $x_c\simeq 0.16$, above which metallic
conductivity sets in at low $T$.
    
In the band model the properties from 1) to 4) are brought about by the
DE- and JT-mechanisms.   We interpret the value
$x_c\simeq 0.16$ in 
terms of percolation theory: both the cluster approach and the phase
separation picture, as suggested first for the cuprates [10], seem to
give the same criterion [11]. 
    
Let us start from the $A$-type of the ground state of
LaMnO$_3$.  According to our picture, appearance of such a
magnetic state is {\it not} caused by any exchange interaction between
localized 
spins on different sites.  The mutual arrangement of the distorted
octahedra and of the 
core $^3t_{2g}$-spins, ${\bf S}_i$ (S=3/2), is stabilized only by gains in the
kinetic {\it band} 
energy of the ``conduction'' $e_{2g}$-electrons.
    
The single electron Hamiltonian is then of the form [6]:
\begin{eqnarray}
\hat{H}=\sum_{i,\delta} \left( \hat{t}_{i,i+\delta}-J_H{\bf
S}_i\cdot\hat{\mbox{\boldmath $\sigma$}}+g\hat{\mbox{\boldmath $\tau$}}_i\cdot{\bf Q}_i+J_{el}{\bf
Q}^2_i\right)
\end{eqnarray}

\noindent Here $\hat{t}_{i,i+\delta}$ is the nearest neighbor tunneling
matrix,  defined on some 
basis of the two-dimensional cubic representation, $e_{2g}$; ${\bf Q}_i$
are the active 
local JT-modes with the matrix elements on that electronic basis in
terms of a ``pseudospin'' matrix,  $\hat{\mbox{\boldmath $\tau$}}$ (see
e.g.,in  [7]; ${\bf Q}_i$ are  
defined as dimensionless parameters).  The static JT deformations and the
staggered magnetization,
$\langle{\bf S}_i\rangle =(-1)^i\langle S\rangle$ of the $A$-type phase
are treated  below in the mean field
approximation.
    
To avoid cumbersome expressions, we first discuss the
competition between ferro- and antiferromagnetic order for the {\it one}
band model: 
\begin{eqnarray}
\varepsilon ({\bf p}) = t\cos (ap_z) + \bar{t} ({\bf p}_\perp)
\end{eqnarray}

\noindent Assume first that the local ($^3t_{2g}$) spins are ordered
ferromagnetically.  At 
large $J_H\gg t$ the low energy band is shifted down, by $-J_H \langle
S\rangle$.  The main
contribution to the total energy is $E_H =-J_H\langle S\rangle n$ ($n$
is the number of 
electrons per unit cell), while the kinetic energy contribution due to the
polarized electrons is linear in $t,\bar{t}$.
     
Consider the same problem for AF-ordering along the $z$-axis (with
the period 2$a$). It is convenient to discuss a more general case:
\begin{eqnarray}
{\bf S}_i =( \pm\langle S_z \rangle, M_x)~;~~ (S^2_z + M_x^2 =S^2)
\end{eqnarray}

\noindent i.e. the canted AF-structure [8] for the $^3t_{2g}$- spins.
Solving the 
periodic  
electron spectrum problem in a manner similar to the solution for
the two local sites [5], one obtains the following four branches:
\begin{eqnarray}
\varepsilon_l ({\bf p})- \bar{t} ({\bf
p}_{\perp})=~~~~~~~~~~~~~~~~~~~~~~~~~~~~\nonumber\\ 
\pm\sqrt{ J_H^2 S^2+
t^2(p_z) \pm 2J_H\mid t(p_z)\mid \mid M_x\mid}
\end{eqnarray}

\noindent At $J_H\gg t,\bar{t}$, the low energy spectrum reduces to:
\begin{eqnarray}
\varepsilon({\bf p})_{1,2} - \bar{t}({\bf p}_{\perp})\simeq
-J_HS\pm\mid t({\bf p}_z)\mid (\mid M_x\mid /S)
\end{eqnarray}

The second term restores the familiar expression for
transport [5]: $M_x= 2S \mid\cos \theta /2\mid$ ($\theta$ is the angle
between the adjacent local spins).
   
For the
AF-case $(M_x\equiv 0)$, the Brillouin zone is reduced by half, but there are
now two branches in (5). As the result, the main term, $-J_HS$, does
not change.  For a single band, the
antiferromagnetic order may even be energetically favorable, although
the energy gain would be small,of the order of (see (4)):
\begin{eqnarray}
t^2/J_H\ll t
\end{eqnarray}                                                  

\noindent For two bands there are terms linear in
$t$.  The JT-band splitting {\it is} necessary if we wish to decrease the
ground state 
energy below that one for the ferromagnetic state. 
   
We  obtained the  ``bare'' band spectrum of the cubic material with the
$e_{2g}$-electrons.  For calculating 
$\hat{t}_{ik}$, and hence, the
electronic spectrum, $\varepsilon ({\bf p})$, use of the normalized
basis functions of the form:
\begin{eqnarray}
\psi_1\propto z^2+\epsilon x^2+\epsilon^2y^2,~ \psi_2\equiv\psi_1^{\ast}
\end{eqnarray}

\noindent ($\epsilon =\exp (2\pi i/3)$), proves to be more
convenient to account 
for the cubic 
symmetry of the initial lattice. The functions (7) are connected with
the real basis, $\varphi_1\propto d_{z^2}$    and
$\varphi_2\propto d_{x^2-y^2}$:
\begin{eqnarray}
\psi_1=(\varphi_1 + i\varphi_2)/\sqrt{2}~;~~
\psi_2=(\varphi_1 - i\varphi_2)/\sqrt{2} 
\eqnum{7'}
\end{eqnarray}

\noindent The tight-binding spectrum of the $e_{2g}$-electrons consists
of two branches, which are discussed below.

Before proceeding further, we write down explicitly the local, JT-part of
the Hamiltonian (1) in the basis (7):
\begin{eqnarray}
-\frac{g}{2}Q_0\left( \begin{array}{cc} 0 & \exp (i\theta) \\ \exp
(-i\theta) & 0 \end{array}\right)
\end{eqnarray}

\noindent In the standard notations [7]:  $Q_2 = Q_0 \sin\theta ~;~
Q_3 = Q_0 \cos\theta$.  Here $Q_0$ 
 is
the magnitude of the JT distortions, while the ``angles'', $\theta =0,
2\pi /3, -2\pi /3$
correspond to elongations of the octahedron along the $z$, $x$, and $y$-axes,
respectively.
   
With the above in mind, consider the options for possible ground
states in manganites.  In the case of stoichiometric LaMnO$_3$ there is exactly
one $e_{2g}$-electron (Mn$^{3+}$) per unit cell.  The ferromagnetic
filling-up for two
bands will produce a metallic spectrum.  AF-ordering along one axis
results in the same main contribution, 
$-J_HS$ (at large $J_H$), while the corresponding kinetic energy
contribution becomes of the
order of  $t^2/J_H\ll t$.  Second order terms in $t$ in Eq. (4) imply
that the  
spectrum is much less dispersive along the z-axis $(M_x\equiv 0)$.
Planes with the antiparallel core spins become almost isolated.
    
AF-ordering along any other direction would lack electronic transport
along this direction, again shrinking bands.  Thus, it is
enough to have the AF order along one axis to access the major energy
gain, $-J_HS$.   However, to 
stabilize the AF-order along {\it one} axis,
one needs to further reduce the total energy by the antiferroelastic
JT-distortions along the remaining directions.
    
Note again that after the AF-ordering sets in, say, along the $z$-axis, the
communication between adjacent layers in real space becomes very weak.
Therefore, at large $J_H$ it is possible to reduce the problem, to main
approximation, to the one of two-dimensional (2D) electrons.  For
decoupled layers, electrons inside the layer are ferromagnetically
polarized along the $^3t_{2g}$-(core) spins.  Again, in this 2D-problem
there is 
one electron per (2D) cell.  The metallic spectrum simplifies to the
form:
\begin{eqnarray}
\varepsilon_{1,2}^{2D}({\bf p})=(A+B)(\cos ap_x+\cos ap_y)\pm \nonumber\\
(A-B)\left\{
(\cos ap_x+\cos ap_y)^2+ \right. \nonumber\\
\left.3(\cos ap_x-\cos ap_y)^2\right\}^{1/2}
\end{eqnarray}

\noindent where $i,k = (x,y,z)~;~~ A\propto \overline{\varphi_1(z;
x,y) \varphi_1(z+a; x,y)}~,~~ B\propto \overline{\varphi_2(z; x,y)\varphi_2
(z+a; x,y)}$ are the two overlap integrals between the two sites
separated by the lattice constant, $a$.
Introduce now JT-distortions in the layer with the structure vector
$(\pi ,\pi)$ in a plane. If
the JT-splitting is strong enough,then instead of (9), the new spectrum
will consist of two pairs of bands separated by energy gaps over the
entire (new) Brillouin zone, with the low-energy bands fully occupied.  This
completes the band picture for the insulating $A$-type ground state in
LaMnO$_3$.
    
For this interpretation one needs:
\begin{eqnarray}
J_H \gg t,~ gQ_0~;~~gQ_0\stackrel{\sim}{>} t
\end{eqnarray}

\noindent Eq.(10) does not impose any severe limitations on the model
parameters.  The above physics should be present in any realistic band
structure. Usually the band calculations use the experimental
lattice parameters, and it is not straightforward to single out
(e.g. in [3]) the competition of the effects considered above.
     
Finding the single electron spectrum in presence of the JT-distortions
from (8,9), writing down the total energy of the filled-up bands and
subsequent minimization of the total energy, is a
straightforward but rather tedious task.  In the general case, there is no
small parameter.  We will consider numerical aspects elsewhere.
     
Below we take advantage of the transparent physics of the
independent 2D-layer approximation.  First of all, the 2D-ferromagnetic
ground state in each layer which was obtained in the mean field
approximation, will be smeared out by fluctuations: ferromagnetism does
not exist as a stable thermodynamical phase in 2D, unless there is a
weak interlayer coupling, $t^2/J_H$, of eq. (6).  This results in a 
rather low N\'{e}el temperature, $T_N\simeq 150K$, in
LaMnO$_3$, at $t\sim 0.1 eV, ~ J_H\sim 1 eV$.

To illustrate the above, consider the ``symmetric'' model [9], $A=B=t/2$
in (9):
\begin{eqnarray}
\varepsilon({\bf p})=t(\cos ap_x+\cos ap_y)\equiv \tilde{t}({\bf p})
\end{eqnarray}

\noindent Eq. (11) gives the two degenerate bands.
For one polarized 2D-electron per unit cell, the two Fermi
surfaces run along the lines $p_x\pm p_y=(\pm \pi /a)$.  In the presence of the
JT-distortion (8) the new spectrum (in the reduced Brillouin zone, with the
structural vector $(\pi /a, \pi /a)$) consists of the four
branches: $\varepsilon_l({\bf p})=\pm\mid \mid t({\bf p})\mid\pm\Delta\mid$,
where $\Delta=\mid gQ_0/2\mid$.  At $\Delta > 2t$ there are two pairs of
energy bands, and the 
insulating gap sets in at $p_x= p_y=0$, equal to $\Delta
-2t$. The 
distortion $gQ_0=g^2/J_{el}$ is defined from the on-site problem.
        
Turning back to the parameters ($J_H, J_{el}, g, A, B$)
of our general model, connections between $gQ_0, J_{el}$ and $A$ and $B$
are established by
minimization of the total (electronic and elastic) energies with respect
to $Q$ and $M_x$ (see eq. (5)). The
deformations of octahedra $(Q_0)$ are known from the
structural data 
(see e.g., in [3]). One may also expect that
$B\ll A$: tunneling between Mn-sites takes places through the shared oxygen
ions, and the notion that the d-shell size, $d$, is small compared to the
lattice constant, $a (d\ll a)$, is quite helpful, as oxides of
transition metals are concerned [7, 13, 14].  Currently the spin wave
spectrum in LaMnO$_3$ is available [15].  Although we postpone the study of the
spin wave spectrum for the future, it is worth mentioning that 
coexistence of the localized $^3t_{2g}$-spins and itinerant
$e_{2g}$-electronic states 
(which in the AF-environment are not characterized by the spin-projection
quantum number), may result in deviations from an anisotropic
Heisenberg model.  

Finally, we briefly discuss the concentration dependence.  At small $x$
the two mechanisms lead to localization of a doped hole.  The first
one is directly related to the 2D physics.  According to
(1), interaction of a hole with the JT-modes may decrease the hole's
energy by an amount of order of $g^2/J_{el}$.  If its
energy is now below the band's bottom, the hole becomes self-trapped.
Two-dimensionality plays the crucial role in that there are no energy
barriers for the process [12].  If a dimensionless parameter $C\sim
g^2/J_{el}W$ ($W$ is a band-width) exceeds a critical value of order of
unity, no itinerant states are possible in the 2D system.  The
second mechanism is the Coulomb binding of holes near dopants [8].  

At larger $x$ the band approach becomes useless.  Instead, we
adopt the 
point of view of the percolation theory. Let us start first with the
{\it cluster} description [11].  In the so-called site problem, consider
formation of an infinite cluster of the neighboring divalent ions.
For the simple cubic 
lattice the critical $x_{cr}(s)=0.31$. This value is not universal.  

It is well known [11] that
correlations between sites rapidly reduce $x_c(s)$ to the value
$x_c\simeq 0.16$, which is also a threshold concentration for a
continuous percolation problem.  Although holes are
located near divalent ions (forming the cluster's skeleton), the hole's
hopping takes place along the adjacent Mn-sites
correlated by the fact that the wave function of a single hole is spread
over a few interatomic distances tending to suppress JT-distortions and
align spins ferromagnetically in its vicinity.  The cluster's structure
is complicated, however, at $x>x_c$ one may expect
that locally its properties are close to the ones of the homogeneous
metallic ferromagnetic state.  The rest of the sample
does not conduct.
    
Another plausible view is that the material may separate into 
coexisting insulator and 
metallic phases [10].  In the cluster approach the Coulomb energy is
playing the crucial role confining holes in the skeleton's vicinity.  In
the continuous description the Coulomb energy would limit sizes of the
phases' domains.  In both approaches
the Coulomb energy may be strongly reduced by sufficiently large
dielectric constant of the insulating phase and due to metallic screening
in the ferromagnetic phase.
     
With the temperature increase, concentration of the metallic phase
decreases: temperature-induced JT-distortions [6] provide
stronger localization of  ``doped'' holes.  From such a point of view
the metal-insulator transition temperature, $T^{\ast}$, may also be treated
as a percolation point (at fixed $x$), and this agrees well with
the extreme sensitivity of CMR to applied magnetic field.

To summarize, the view of LaMnO$_3$ 
as a band insulator, is consistent with the main experimental facts.
Large intraatomic Hunds' exchange, $J_H$, forces the cubic 
system to order antiferromagnetically along one direction because the
band energy of 
electrons can then be effectively reduced by the JT-instability
resulting in the new 
(tetragonal) lattice ($\sqrt{2}a\times\sqrt{2}a\times 2a$). At $J_H\gg
t,gQ_0$, the 
$e_{2g}$-electrons  form
almost disconnected 2D-layers.  This manifests itself in
the low N\'{e}el temperature.  The percolation
analysis can explain the value of the critical concentration, $x_c\simeq
0.16$ and CMR itself suggesting new interesting physics.
     
The work was supported (L.P.G.)by the National High Magnetic Field
Laboratory through NSF Cooperative Agreement \# DMR-9016241 and the State of
Florida, and (V.Z.K.) by the U.S. Office of Naval research under contract
\# N00014-97-F0006.

\vspace{.25in}

\noindent [1] H. Kawano et al., Phys. Rev. B \underline{53}, 14709 (1996).

\noindent [2] J. Coey, M. Vivet, and S. von Molnar, (unpublished, \\
\indent $~$ 1997).

\noindent [3] W. Pickett and S. Singh, Phys. Rev. B \underline{53},
1146 \\ \indent $~$ (1996). 

\noindent [4] G. Zener, Phys. Rev. \underline{82}, 403 (1951).

\noindent [5] P.W. Anderson and H. Hasegawa, Phys. Rev. \underline{100},
\\ \indent $~$ 675 (1955). 

\noindent [6] A. Millis, B. Shraiman, and R. Mueller, Phys. Rev. 
\\ \indent $~$ Lett. \underline{77}, 175 (1996).

\noindent [7] J. Kanamori, J. Appl. Phys. (Suppl.) \underline{31}, 145 (1960).

\noindent [8] P.G. de Gennes, Phys. Rev. \underline{118}, 141 (1960).

\noindent [9] For review, see K. Kugel and D. Khomskii, Sov.Phys. \\
\indent $~$ - Usp.
\underline{25}, 231 (1982).

\noindent [10] L.P. Gor'kov and A. Sokol, JETP Lett., \underline{46},
420 \\  \indent $~~$ (1987).

\noindent [11] H. Scher and R. Zallen, J. Chem. Phys. \underline{53},
3759 \\  \indent $~~$ (1970);for review see
B. Shklovskii and A. Efros, \\ \indent $~~$ {\it Electronic
Properties of Doped 
Semiconductors} \\ \indent $~~$ (Springer -Verlag, Berlin, 1984); G. Deutcher,
in  \\ \indent $~~$ {\it Chance 
and Matter}, p. 1, edited by J. Souletie, J.  \\ \indent $~~$ Vannimenus, 
and R. Stora (Elsevier, Amsterdam,  \\ \indent $~~$ 1987).

\noindent [12] E. Rashba, in {\it Excitons}, p. 473, edited by E. Rashba
\\ \indent $~~$ and M. Sturge,  (North-Holland, Amsterdam, 1982); \\
\indent $~~$ Y.
Toyozawa,  Y. 
Shinozuka, J. Phys. Soc. Jpn. \underline{52}, \\ \indent $~~$ 1446 (1983).

\noindent [13] J. Coodenough, in {\it Progress in Solid State
Chemistry}, \\ \indent $~~$ v. 5, p. 145, edited by H. Reiss (Pergamon,
\\ \indent $~~$ NY, 1971). 

\noindent [14] P.W. Anderson, in {\it Magnetism}, v. 1, p. 75, edited \\
\indent $~~$ by
G. Rado and H. Suhl (Academic, NY, 1963).

\noindent [15] K. Hirota, N. Kaneko, A. Nishizwa, and Y. Endoh, \\
\indent $~~$ J.
Phys. Soc. Jpn. {\bf 65}, 3763 (1996).

\end{document}